\documentclass[prd, amsfonts, twocolumn, nofootinbib, showpacs]{revtex4}
\usepackage{graphicx}
\usepackage{epsfig}
\usepackage{color}
\usepackage{amsmath}
\usepackage{amssymb}
\newcommand{\be}{\begin{equation}}
\newcommand{\ee}{\end{equation}}
\newcommand{\bea}{\begin{eqnarray}}
\newcommand{\eea}{\end{eqnarray}}

\newcommand{\gapp}{\mathrel{\raise.3ex\hbox{$>$}\mkern-14mu
\lower0.6ex\hbox{$\sim$}}}
\newcommand{\lapp}{\mathrel{\raise.3ex\hbox{$<$}\mkern-14mu
\lower0.6ex\hbox{$\sim$}}}
\def\bbox{{\,\lower0.9pt\vbox{\hrule \hbox{\vrule height 0.2 cm
\hskip 0.2 cm \vrule  height 0.2 cm}\hrule}\,}}

\begin{document}
\title{Strong lensing constraints on modified gravity models}
\author{De-Chang Dai$^{1,3}$\footnote{corresponding author: De-Chang Dai,\\ email: diedachung@gmail.com\label{fnlabel}}, Dejan Stojkovic$^2$, Glenn D. Starkman$^3$}
\affiliation{$^1$ Center for Gravity and Cosmology, School of Physics Science and Technology, Yangzhou University, 180 Siwangting Road, Yangzhou City, Jiangsu Province, P.R. China 225002}
\affiliation{ $^2$ HEPCOS, Department of Physics, SUNY at Buffalo, Buffalo, NY 14260-1500}
\affiliation{ $^3$  CERCA/Department of Physics/ISO, Case Western Reserve University, Cleveland OH 44106-7079}
 %%%%%%%%%%%%%%%%%%%%%%%%%%%%%%%%%%%%%%%%%%%%%%%%%%%%%%%

\begin{abstract}
\widetext
We impose the first strong-lensing constraints on a wide class of modified gravity models where an extra field that modifies gravity also couples to photons (either directly or indirectly through a coupling with baryons) and thus modifies lensing.
We use the nonsingular isothermal ellipsoid (NIE) profile as an effective potential which produces flat galactic rotation curves.
If a concrete modified gravity model gives a flat rotation curve, then the parameter $\Gamma$ that characterizes the lensing effect must take some definite value. We find that $\Gamma=1.24\pm 0.65$ %$\Gamma=1.24275\pm 0.64866$
at $1\sigma$, consistent with general relativity ($\Gamma=1$).
This constrains the parameter space in some recently proposed models.
\end{abstract}

%%%%%%%%%%%%%%%%%%%%%%%%%%%%%%%%%%%%%%%%%%%%%%%%%%

\maketitle

Astrophysical observations on scales greater than the solar system do not match predictions from standard gravity sourced by ordinary matter. On the scales of galaxies and clusters of galaxies the problem might be resolved if we postulate the existence of particle dark matter. However, the persistent null results from direct and indirect searches for particle dark matter strengthen the case to seriously explore alternative explanations.  Those include alternative dark matter candidates such as primordial black holes and macros, but also the possibility that General Relativity must be altered.

The principal alternative gravity framework, MOND, has been around for decades as a phenomenological fit\cite{Milgrom:1983ca}. More recently concrete dynamical models with well defined relativistic Lagrangians have emerged\cite{Bekenstein:2004ne}. By construction, such models reproduce flat galactic rotation curves, however, as hinted in \cite{Berezhiani:2015bqa}, so far it has not been considered whether these models can pass the strong lensing test on galactic scales.

General relativity has been tested to high precision on the solar system scale. Recently such tests have been also extended to galactic scales \cite{Bolton:2006yz,2010ApJ...708..750S,Cao:2017nnq,Collett:2018gpf}. These tests mainly focus on testing the post Newtonian parameters, in particular $\gamma$, which is the leading order term. The geometric metric in the post-post-Newtonian form
can be written as
\begin{equation}
\label{dssqpostpost}
d\tilde{s}^2=-\Bigg(1+2U(x_i,t) \Bigg)dt^2+\Bigg(1-2\gamma U(x_i,t) \Bigg)dx_i^2 ,
\end{equation}
where we have included only the leading-order terms.
The Newtonian potential is given by
\begin{equation}
U(x_i,t)=-\int \frac{\rho(\vec{x'},t)}{|\vec{x}-\vec{x'}|}d^3 x' \,,
\end{equation}
where $\rho$ is the total mass density.
This includes both regular and dark matter, if any.

In GR (with or without dark matter), a single metric determines the motion of both matter and photons. For some modified gravity models this is not the case.
Here, we focus on  modified gravity models that include an extra field that couples to the ordinary matter to explain its motion without dark matter. Since this extra field couples to baryons, it must couple at least indirectly to photons too. Most often, a direct coupling of this additional field to photons is not included \cite{Milgrom:1983ca,Bekenstein:2004ne,Verlinde:2016toy,Dai:2017qkz,Hossenfelder:2017eoh,Dai:2017guq,Edmonds:2017fce}, or if it is included the magnitude of the coupling is not specified \cite{Berezhiani:2015bqa}. Such a non-geometric effect on the photon's path would modify the galactic gravitational lensing signature, and could directly provide a constraint on these models.

Consider such a new field, $\phi_M$, which is introduced to reproduce galactic rotation curves without dark matter.
In this case, the effective metric (see e.g. \cite{Berezhiani:2015bqa}) may be written in the form
\begin{equation} \label{mgl}
d\tilde{s}^2=-\left[1+2\left(\Phi_b + \phi_M\right) \right]dt^2+\left[1-2\left(\Phi_b +\Gamma \phi_M\right) \right]dx_i^2 .
\end{equation}
$\Phi_b$ is the gravitational potential from ordinary matter,
which is insufficient to produce flat galactic rotation curves. 
In general, $\phi_M$ does not have to couple to photons
in the same way as in Einstein's gravity ($\Gamma =1$),
and thus can bend light differently. 

Two things are apparent from comparing the metrics
\eqref{dssqpostpost} and \eqref{mgl}:
first, $\Gamma $ is not equivalent to $ \gamma$;
and second, one cannot test this class of models
just using the $g_{00}$ component of the metric.
So far $\Gamma$ has not been constrained,
and theoretically any value
(even $\Gamma=-1$ as pointed out in \cite{Berezhiani:2015bqa})
is possible.
The main purpose of this paper is to constrain  $\Gamma$.

In our analysis we assume that the extra contribution from $\phi_M$
must produce a flat rotation curve at large galactic radii.
In order to reproduce this behavior
and to also construct the galactic bulge and disk,
we use a nonsingular isothermal ellipsoid (NIE) profile.
Following \cite{2011MNRAS.417.1621D},
a NIE profile in a cylindrical coordinate system
can be parameterised as \cite{1998ApJ...495..157K}
\begin{equation}
	\rho_{NIE}(R,z,V_c,r_c,q_3)=\frac{V_c^2}{4\pi G q_3}\tabularnewline
		\frac{e/\sin^{-1}e}{r_c^2 + R^2 +z^2/q_3^2}\,.
\end{equation}
Here, $V_c$ is the asymptotic circular velocity,
$r_c$ is the core radius,
$q_3$ is the three-dimensional axis ratio,
and $e = \sqrt{1- q^2_3}$.
The circular velocity  profile is then \cite{1998ApJ...495..157K}
\begin{eqnarray}
\frac{V_{NIE}^2(R,V_c,r_c,q_3)}{V_c^2}=&& \\
&&\!\!\!\!\!\!\!\!\!\!\!\!\!\!\!\!\!\!\!\!\!\!\!\!\!\!\!\!\!\!\!\!\!\!\!
1-\frac{e}{\sin^{-1}e}\frac{r_c}{\sqrt{R^2+e^2r_c^2}}\tan^{-1}\Big(\frac{\sqrt{R^2+e^2r_c^2}}{q_3r_c}\Big) \,.\nonumber
\end{eqnarray}

Meanwhile, gravitational lensing causes
the distortion of the image of a background source.
The deflection angles in the lensed image
are given by \cite{1998ApJ...495..157K}
\begin{eqnarray}
	\label{bendingangles}
	\alpha_x &=&F\frac{b}{\sqrt{1- q^2}}\tan^{-1}\Big(\frac{x\sqrt{1-q^2}}{\Psi+r_c}\Big)\\
	\alpha_y &=&F\frac{b}{\sqrt{1- q^2}}\tanh^{-1}\Big(\frac{y\sqrt{1-q^2}}{\Psi+q^2r_c}\Big)\,,\nonumber
\end{eqnarray}
 where $\Psi^2 = q^2 (r_c^2+x^2)+y^2$,
 $b=b_{sis}(e/\sin^{-1}e)$,
 $b_{sis}=2\pi (V_c/c)^2D_{ds}/D_s$.
 Here $D_d$ is the distance from the observer
 	(us) to the gravitational lens,
 $D_s$ is the distance from us to the source that is lensed,
 and $D_{ds}$ is the distance from the lens to the source.
 The parameter $q$ is the  axis ratio
 	of the projected mass distribution
\begin{equation}
q = (q^2_3\sin^2i + \cos^2 i )^{1/2} .
\end{equation}

The key quantity in \eqref{bendingangles} is $F$,
which depends on the source of the lensing
%\begin{eqnarray}
%F&=& 1 \textbix{ , for regular matter}\\
%\end{eqnarray}
\begin{equation}
	F=
	\begin{cases}
	1 &\text{, for regular matter}\\
	\frac{1+\Gamma}{2} &\text{, for the extra field $\phi_M$} .
	\end{cases}
\end{equation}

The NIE profile must reproduce three galactic components:
the extra field,  the bulge and the disk.
The extra field $\phi_M$ is responsible for the asymptotic flat rotation curve,
therefore at large radii we can approximately ignore the contribution
from the matter and take
\begin{eqnarray}
\phi_M(\vec{x},V_{cM},r_{cM},q_{3M})&=&\\
	&&\!\!\!\!\!\!\!\!\!\!\!\!\!\!\!\!\!\!\!\!\!\!\!\!\!\!\!\!\!\!\!\!\!\!\!\!
	\int \frac{\rho_{NIE}(R',z',V_{cM},r_{cM},q_{3M})}{|\vec{x}-\vec{x'}|}d^3 x' \,.\nonumber
\end{eqnarray}

% The asymptotic circular velocity $V_{cM}$ is constant.
% In MOND, $V_{cM}$ is related to baryon density
% 	through the acceleration parameter $a_0$.
% {\color{red} Glenn Why is this MOND statement here?}
We adopt a broad Gaussian prior for $V_{cM}$
%as $N(280,50^2)$, where $N(a,b^2)$ is a Gaussian distribution
with central value $280\text{km/s}$
	and standard deviation $50\text{km/s}$.
	(We write $N(280,50^2)$).
This is meant to be an uninformative, but not indifferent, prior,
reflecting the broad range of observed asymptotic circular velocities.
Similarly, we adopt a uniform prior for $r_{cM}$
	between $0.01\text{arsec}$ and $10\text{arsec}$
	(which we write as $U(0.01,10)$);
%	The prior distribution for $r_{cM}/\text{arsec}$ is $U(0.01, 10)$. $U(a, b)$ denotes a uniform distribution with lower and upper limits, $a$ and $b$, respectively.
and a lognormal prior for $q_{3M}$, with central value $1$
	and standard deviation for $\log(q_{3M})$ of $0.3$.
% The prior distribution for $q_{3M}$ is $LN (1, 0.3^2)$. $LN (a, b^2)$ denotes a lognormal distribution, where $a$ is the central value
% of the variable, and $b$ is the standard deviation for the log of the variable.

The stellar disk and bulge mass distributions
	are assumed to each have S{\`e}rsic profiles,
	which can be approximated with a chameleon profile \cite{Maller:1999de}
\begin{eqnarray}
&&\rho_{Chm}(R,z,V_c,r_c,q_3,\alpha)=\\
&&\quad \rho_{NIE}(R,z,V_c,r_c,q_3)
					-\rho_{NIE}(R,z,V_c,r_c/\alpha,q_3)\, .\nonumber
\end{eqnarray}
(The exact transformation between them can be found in
	the appendix of \cite{2011MNRAS.417.1621D}.)
	
%The disk and bulge's density profile's respective variables are listed in table \ref{parameter}.
The circular velocity due to a chameleon profile is
\begin{eqnarray}
&&V_{chm}^2(R,V_c,r_c,q_3,\alpha)=\\
	&&\quad V_{NIE}^2(R,V_c,r_c,q_3)-V_{NIE}^2(R,V_c,r_c/\alpha,q_3)\, .\nonumber
\end{eqnarray}

The total circular velocity from the different contributions simply add:
\begin{equation}
V^2(R)=V_{bulge}^2(R)+V_{disk}^2(R)+V_{cM}^2(R)\,.
\end{equation}
Likewise, the deflection angles due to a chameleon profile
are given by the differences between the deflection angles
due to each of its NIE components,
and the deflection angles of the bulge, disk and $\phi_M$ are simply the sum
of the three contributions.

We consider specifically SDSS J2141-0001 as the lens galaxy.
The density profile of SDSS J2141-0001
	has already been studied \cite{2011MNRAS.417.1621D,2012MNRAS.423.1073B}.
Here we focus on constraining the parameter $\Gamma$ in modified gravity,
from both gravitational lensing and the galaxy's rotation curve.
We use the lensing data from Hubble Space Telescope (HST)
observations in the F450W (4400s), F606W (1600s) and F814W(420s) filters.
For  SDSS J2141-0001,
$D_d = 497.6\text{Mpc}$,
$D_s = 1510.2\text{Mpc}$, and $D_{ds} = 1179.6\text{Mpc}$.

The lens galaxy surface density profile
is considered to be a combination of two
S\'{e}rsic profile components \cite{2011MNRAS.417.1621D}
\begin{eqnarray}
\Sigma(x,y)=\Sigma_0 \exp\Bigg(-\Big(\frac{R}{R_0}\Big)^n\Bigg) ,
\end{eqnarray}
representing the disk and bulge.
Here $R=\sqrt{x^2+y^2/q^2}$.
The S\'{e}rsic index $n=1$ for the disk,
and $n=1.21\pm 0.11$ for the bulge \cite{2011MNRAS.417.1621D}.
The bulge major-axis half-light radius and axis ratio
are $R_{50}^b = 0.26 \pm 0.01''$ and $q_b = 0.53\pm 0.02$,
while the disk major axis half-light radius
and axis ratio are $R_{50}^d = 2.53\pm 0.13''$ and $q_d = 0.31\pm 0.02$.
The disk inclination is $q = 0.20 \pm  0.02$
(found measuring the axis ratio of the star-forming ring),
which gives the inclination angle $i = 78.5 \pm 1.2^{\circ}$.
The 3D minor-to-major axis ratio, $q_3$, is given by
\begin{equation}
q^2_3 = (q^2- \cos^2 i)/(1 - \cos^2 i)
\end{equation}

Based on the values of these parameters,
we use the priors provided by \cite{2011MNRAS.417.1621D}.
The bulge 2D axis ratio, $q_{\text{bulge}}$,
	has a lognormal prior distribution
	$LN(0.53,0.03^2)$.
	%with central value $0.53$ and standard deviation $0.03^2$.
The bulge chameleon size, $R_{0,\text{bulge}}$,
	has a lognormal prior distribution
	$LN(0.094,0.03^2)$.
	%with central value $0.094^{\prime\prime}$
	%and standard deviation $(0.03^{\prime\prime})^2$.
The bulge chameleon index, $\alpha_{\text{bulge}}$, is $0.4892$.
The disk 2D axis ratio, $q_{\text{disk}}$,
	has a lognormal prior distribution
	%with central value $0.31$
	%and standard deviation $(0.03)^2$.
	$LN(0.31,0.03^2)$.
The disk chameleon size, $R_{0,\text{disk}}/\text{arsec}$,
	also has a lognormal prior distribution
	%with central value $1.10$ and standard deviation $0.03^2$.
	$LN(1.10,0.03^2)$.
The disk chameleon index, $\alpha_{\text{disk}}$, is $0.63$.
The cosine of disk inclination angle, $\cos(i)$, is $0.2$.

\begin{figure}
   %\centering
\includegraphics[width=10cm]{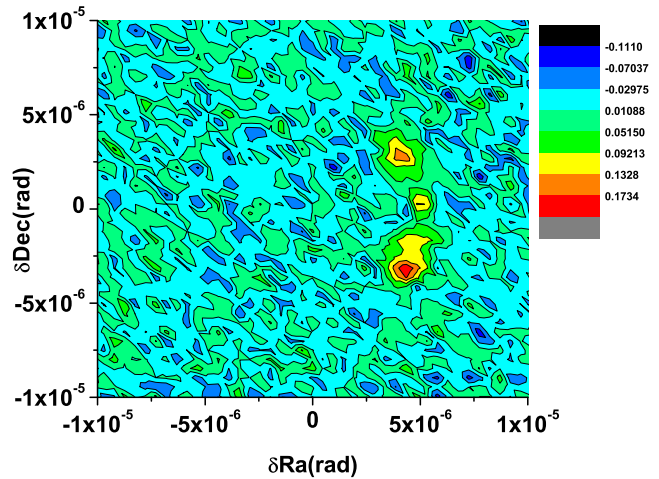}
\caption{This is the lensing image of SDSS J2141-0001 from the filter 606W. The lens galaxy is removed. The yellow arch is the distorted source galaxy.
}
\label{map}
\end{figure}

To study the pure gravitational lensing effect,
we must keep the strongly lensed galaxy
and remove the lensing galaxy and other non-related light sources.
In this we follow the method from \cite{2011MNRAS.417.1621D}.
Galactic light around the arc is subtracted
by reflecting the galaxy along the minor axis.
Then the arc is marked and noise is added
	at the level of $\sigma_{15} = 15\%$ of the peak arc brightness.
This $15\%$ value is an estimate from a Poisson noise distribution.
The final image is shown in figure \ref{map}.
We see that an arch structure appears, which is the lensing source galaxy.

So far we dealt only with the shape parameters.
The total initial mass is obtained
from a Chabrier initial mass function \cite{Chabrier:2003ki}.
The prior for the bulge initial mass, $\log_{10}(M_{*,b}/M_\odot ) $,
is $N(10.26, 0.08^2)$;
while the prior for the disk initial mass, $\log_{10}(M_{*,d}/M_\odot)$,
is $ N(10.88,  0.07^2)$ \cite{2011MNRAS.417.1621D}.
We use this initial function for most of our analysis.
We also include one case with the same initial mass prior
	as in \cite{2011MNRAS.417.1621D} --
	a prior for the stellar mass, $\log_{10}(M_*/M_{\odot})$, of $U(10.5,11.4)$ and a prior for the  bulge stellar mass fraction, $f_{bulge}$,
	of $N(0.2,0.04^2)$.

Since the galactic center-of-mass
may be offset from the galaxy's center-of-light,
a prior distribution for the offset is also included.
The priors for the spatial offset in the $x$  direction, $x_c/\text{arcsec}$,
and in the  $y$ direction, $y_c/\text{arcsec}$, are each taken to be
$N (0, 0.01^2)$.
The prior on the mass-light position angle offset, $\theta/\text{deg}$,
 is $N (1.7, 2.9^2)$.

The external shear may also cause distortion,
so we include it in one of the cases.
The prior on the lens external shear, $\gamma_g$,  is  $N (0, 0.1^2)$.
The prior on the position angle of external shear, $\theta_{\gamma}/\text{deg}$,
is $U(0, 180)$.

 We use the rotation curve data provided by \cite{2011MNRAS.417.1621D},
 which was observed with DEIMOS on Keck II on October 1st 2008.
 The grating slit width is $ 1"$.
 It covers a range of radii,
 therefore beam-smearing effect must be included \cite{2011MNRAS.417.1621D}.

In our case, the likelihood curve
includes both the rotation curve and the gravitational lensing.
We take the rotational-curve data,
velocities ($v(r_j)$) and standard deviations ($\sigma_j$)
from \cite{2011MNRAS.417.1621D}.
The likelihood for the rotation curve is
\begin{equation}
Pr(v|\theta_m) \propto  \exp(-\frac{\chi_v^2}{2})
\end{equation}
with
\begin{equation}
\chi^2_v=\sum_j \frac{(v(r_j)-v_j^p)^2}{\sigma_j^2} .
\end{equation}
$v_j^p$ is the velocity at $r_j$ from the prior.
Since the slit width is not small, the beam-smearing effect must be included. The  likelihood for the lensing is
\begin{equation}
Pr(d|\theta_m,\theta_s) \propto \exp(-\frac{\chi_d^2}{2T})
\end{equation}
with
\begin{equation}
\chi^2_d=\sum_j \frac{(d(\vec{r}_j)-d_j^p)^2}{2\sigma_{15}^2} ,
\end{equation}
where $d(\vec{r}_j)$ is the lensing-image pixel value at $\vec{r}_j$,
while $d_j^p$ is the predicted pixel value at $\vec{r}_j$.
We assume the source also has
a S{\'e}rsic profile\cite{2011MNRAS.417.1621D,Marshall:2007tf}.
$T$ is a parameter that tunes the noise level,
which is included because the error cannot be reliably estimated in another way.
 However, if we assume that the noise satisfies the Poisson noise distribution, $T$ ranges from $0.5$ to $1$.
 
The total probability is the combination of
rotation-curve and lensing likelihoods
\begin{equation}
Pr(v,d|\theta_m,\theta_s)=Pr(v|\theta_m)\times Pr(d|\theta_m,\theta_s) .
\end{equation}

We marginalize over all other parameters and leave only $\Gamma$.

\begin{figure}
   %\centering
\includegraphics[width=10cm]{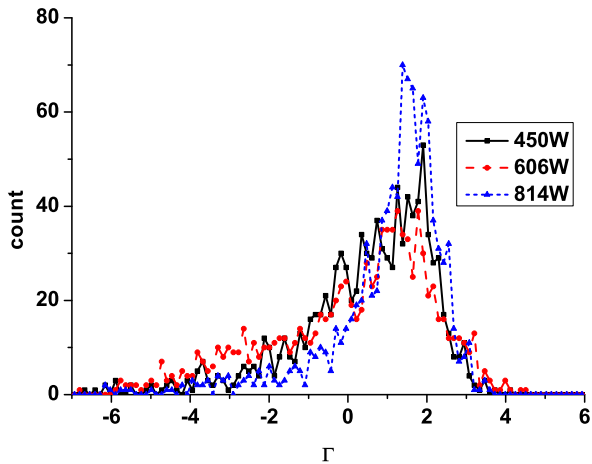}
\caption{The black, red and blue curves
are the probability distributions
for the parameter $\Gamma$ from 450W, 606W and 814W filters.
We set $T =  2$ in all three cases.
The image of the 814W filter has better resolution,
so its distribution is much more concentrated than the other two.
}
\label{450-606-814}
\end{figure}

\begin{figure}
   %\centering
\includegraphics[width=10cm]{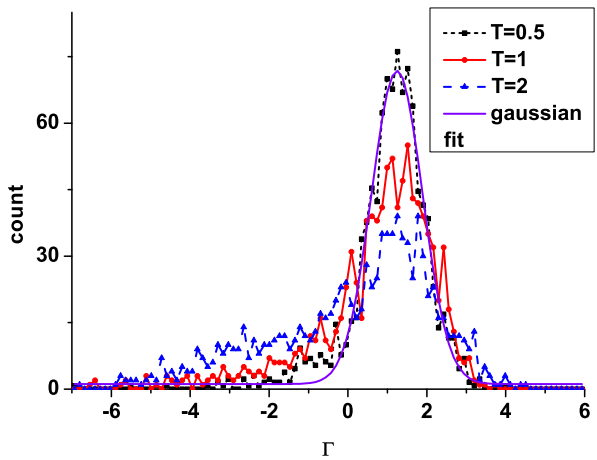}
\caption{The black, red and blue curves
are the probability distributions for the parameter $\Gamma$
for $T=2$, $T=1$ and  $T=0.5$.
The smallest $T$ value is more concentrated than the larger $T$ values.
The violet curve is a Gaussian fit for the $T=0.5$ case.
We find $\Gamma=1.24\pm 0.65$ at $1\sigma$.
}
\label{606}
\end{figure}

In our analysis,
the prior for $\Gamma$ is assumed to be $U(-7,6)$.
Figure \ref{450-606-814} shows the marginal distribution of $\Gamma$
from different filters.
$T$ is chosen to be $2$ in all the cases.
As expected, the highest possible $\Gamma$ is not very far away from $1$.
It can be seen that negative $\Gamma$ is also possible.
However, if we reduce $T$, the probability for negative $\Gamma$ declines,
and the distribution sharpens.
This can be seen in figure \ref{606}.
The $\Gamma$ distribution can be found
by fitting the marginal distribution for $T=0.5$.
At the $1\sigma$ level, we have $\Gamma=1.24\pm 0.65$.
Apparently, $\Gamma<0$ is disfavored,
while $\Gamma=1$ is well within the $1\sigma$ region.

\begin{figure}
   %\centering
\includegraphics[width=10cm]{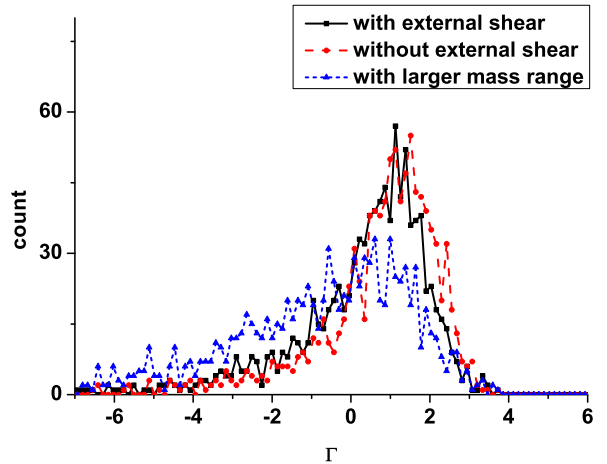}
\caption{The black, red and blue curves
are the probability distributions for $\Gamma$
with the external shear, without the external shear,
and for a larger mass range respectively.
The value for $T$ is chosen to be $1$ in all cases.
The larger mass range case has a prior on
the stellar mass $\log_{10}(M_*/M_{\odot})$ of $U(10.5,11.4)$,
and a prior on the bulge stellar mass fraction $f_{bulge}$ of $N(0.2,0.04^2)$.
The other two use  Chabrier initial mass distributions.
}
\label{606-compare}
\end{figure}

Optionally, we may also include the external shear
and vary the initial mass function distribution.
Figure \ref{606-compare} shows that including external shear
does not change the distribution significantly.
But if the mass range becomes broader,
the $\Gamma$ distribution also becomes broader.

To summarize, using galactic strong lensing data, we imposed constraints on a wide class of modified gravity models where an extra field that modifies gravity also couples to photons (either directly or indirectly through a coupling with baryons) and thus modifies lensing.
We find that the modified gravity parameter $\Gamma$ marginal distribution is $\Gamma=1.24\pm 0.65$.
Therefore it is unlikely to have a negative $\Gamma$.

This constraint applies to most asymptotic circular velocity models.
For example, in TeVes \cite{Milgrom:1983ca,Bekenstein:2004ne} the effective $\Gamma$ is $1$, so this model is consistent with the strong lensing data.
For the superfluid model in \cite{Berezhiani:2015bqa} the effective $\Gamma$ can take large range of values (even negative), so more precise model building would have to take this constraint into account.
Models where there is no (direct or indirect) coupling between the photons and extra degrees of freedom, and thus effectively yield  $\Gamma = 0$ seem to  be excluded at this confidence level. This includes the minimal MOND model, Verlinde's original model \cite{Verlinde:2016toy,Dai:2017qkz} and the version in \cite{Hossenfelder:2017eoh,Dai:2017guq}, and the model in \cite{Edmonds:2017fce}. Appropriate extensions of these models can perhaps be made consistent with the strong lensing data.

The limit on $\Gamma$ we present here is weaker than existing limits
on the usual post-Newtonian parameter $\gamma$. The reason is that $\Gamma$ is sourced only
by extra degrees of freedom whose  contribution to lensing distortion is smaller than from the regular matter.

To obtain this limit,
we  assumed that the contribution
from the extra field coming from modification of gravity
can be approximated as a NIE profile,
without dealing with a specific modified gravity model.
If one uses a specific model, a better constraint may be obtained.

The constraints may also be improved with higher quality data
(e.g. Keck vs HST \cite{2011MNRAS.417.1621D}). The other way to extend the analysis is to include more galaxies.
If it is found that different galaxies prefer different values of $\Gamma$,
this will be very difficult to accommodate with simple modified gravity models.  This would indicate that the empirical flat rotational curve fails somewhere, and cannot be a good benchmark for building a modified gravity model in question.

\begin{acknowledgments}
D.C Dai was supported by the National Science Foundation of China (Grant No. 11433001 and 11775140), National Basic Research Program of China (973 Program 2015CB857001) and  the Program of Shanghai Academic/Technology Research Leader under Grant No. 16XD1401600.
GDS was supported in part by grant DE-SC0009946 from the US DOE
to the particle-astrophysics theory group at CWRU. 
DS was partially supported by the NSF grant PHY 1820738.
\end{acknowledgments}

\end{document}